\documentclass[3p,times]{elsarticle}

\usepackage{ecrc}

\usepackage{epstopdf}

\volume{00}

\firstpage{1}

\journalname{Nuclear Physics A}

\runauth{Author1 et al.}

\jid{nupha}

\jnltitlelogo{Nuclear Physics A}

\usepackage{graphicx}
\usepackage{amsmath,amssymb}

\begin{document}

\begin{frontmatter}



\title{The fate of the weakly-bound $\psi(2s)$ in nuclear matter}

\author{J. Matthew Durham (for the PHENIX\fnref{col1} Collaboration)}
\runauth{J. Matthew Durham (for the PHENIX\fnref{col1} Collaboration)}
\fntext[col1] {A list of members of the PHENIX Collaboration and acknowledgements can be found at the end of this issue.}
\address{Los Alamos National Laboratory, Los Alamos, NM, USA 87545}



\begin{abstract}
We present new results of a completed PHENIX analysis of $\psi(2s)$ modification at midrapidity in 200 GeV $d+$Au collisions. Strong suppression of the $\psi(2s)$ relative to the $J/\psi$ is observed.  This difference in suppression is too strong to be explained by breakup effects in the nucleus, due to the short nuclear crossing times at RHIC. Given the observation of long range correlations in $p(d)+$A collisions at LHC and RHIC, consistent with hydrodynamics, these observations raise interesting questions about the mechanism of $\psi(2s)$ suppression when it is produced in a nuclear target.

In 2012, the PHENIX Collaboration installed the FVTX, a silicon tracker that precisely measures muon pair opening angles prior to any multiple scattering in the muon arm absorber, and thus provides an improved dimuon mass resolution.  The FVTX allows the $\psi(2s)$ to be separated from the $J/\psi$ at forward and backward rapidity for the first time at RHIC.  We present new results on $\psi(2s)$ production in $p+p$ collisions at $\sqrt{s} = 510$ GeV from the 2013 data set.
\end{abstract}


\end{frontmatter}



\section{Introduction}

Charmonia production in relativistic nuclear collisions has been a subject of intense theoretical and experimental investigation for decades (see \cite{Jpsi_theory_review} for a recent review).  While far from the unambiguous signature of deconfined plasma formation expected from early calculations \cite{MatsuiSatz}, charmonia measurements have nevertheless been a source of considerable insight into the dynamics of particle production in hot and cold nuclear matter.  The PHENIX  experiment was specifically designed to measure charmonia production through decays to dielectrons at midrapidity and dimuons at forward rapidity \cite{PHENIXNIM}.  In these proceedings, we present finalized PHENIX results on $\psi(2s)$ production in $d+$Au collisions at midrapidity \cite{PPG151}, and show the first results from a new silicon tracking detector, the FVTX, which enables the $\psi(2s)$ to be measured at forward rapidity for the first time at RHIC.

\section{Charmonia suppression in $d+$Au collisions at RHIC }

\begin{figure}
\centering
\begin{minipage}{.29\textwidth}
\centering
\includegraphics[width=\linewidth]{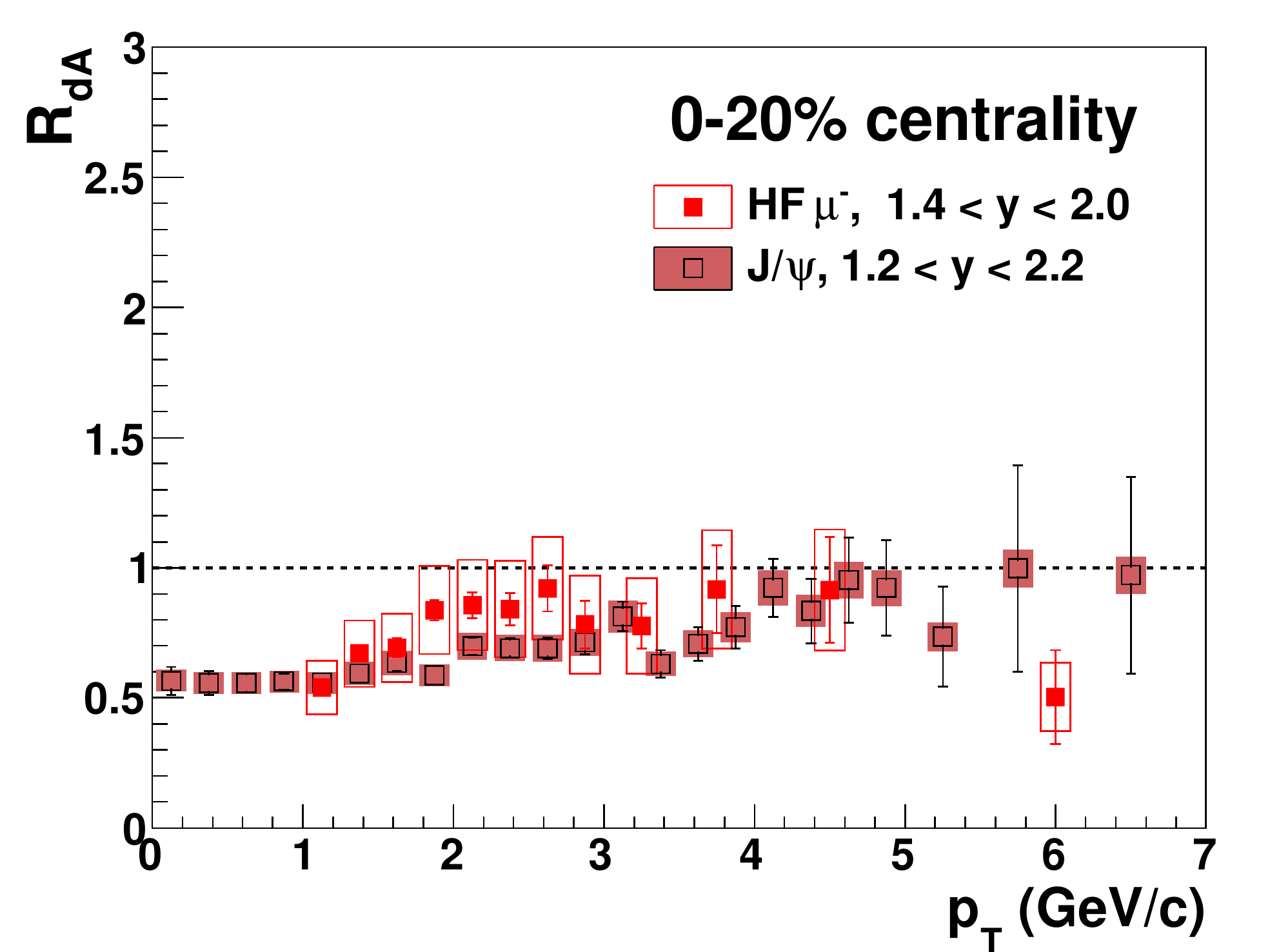}
\caption{$R_{dA}$ of open and hidden heavy flavor at forward rapidity ($d-$going direction).}
\label{fig:fwd}
\end{minipage}\hfill
\begin{minipage}{.33\textwidth}
\centering
\includegraphics[trim = 0mm 0mm 0mm 12mm, clip, width=\linewidth]{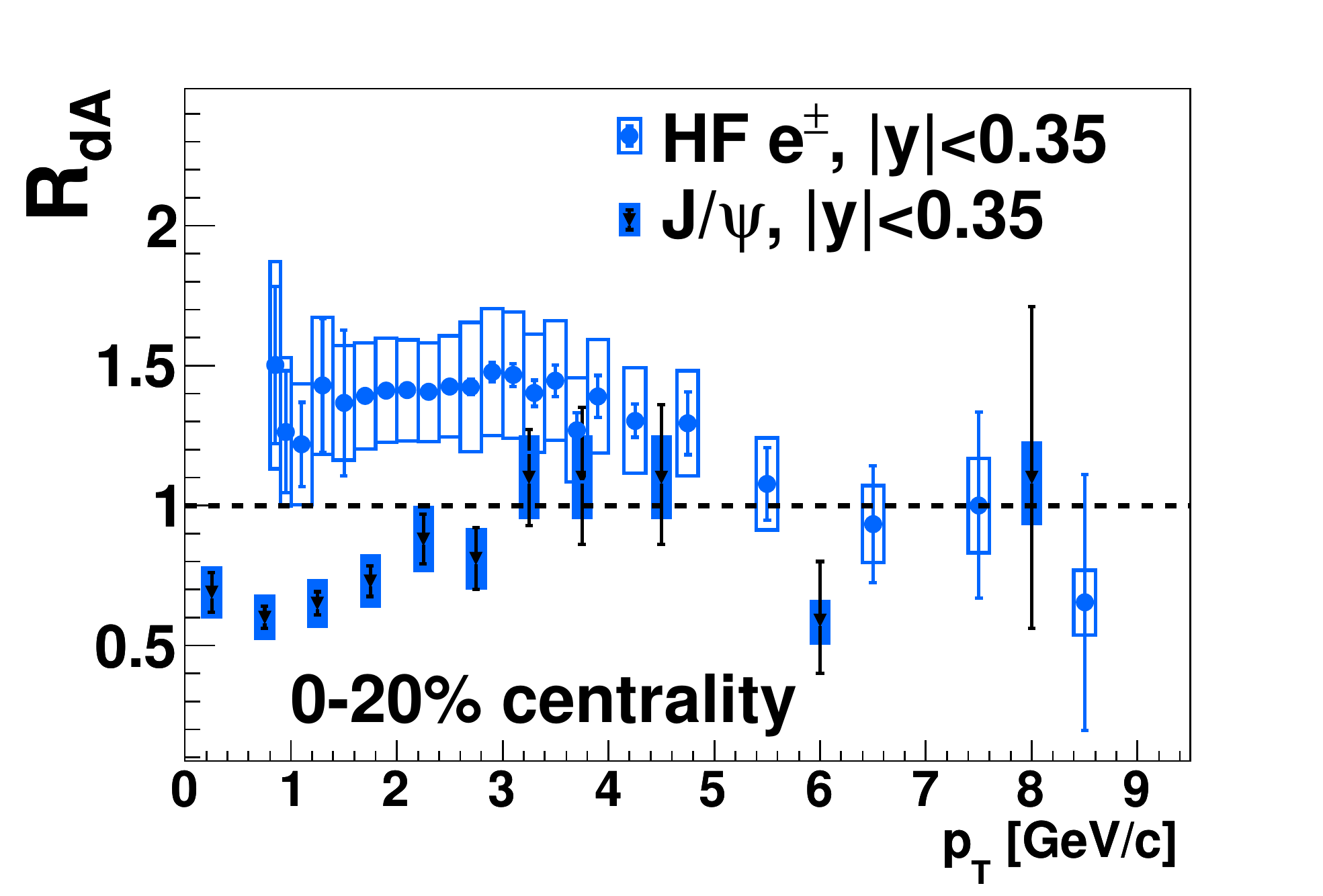}
\caption{$R_{dA}$ of open and hidden heavy flavor at midrapidity.}
\label{fig:mid}
\end{minipage}\hfill
\begin{minipage}{.29\textwidth}
\centering
\includegraphics[width=\linewidth]{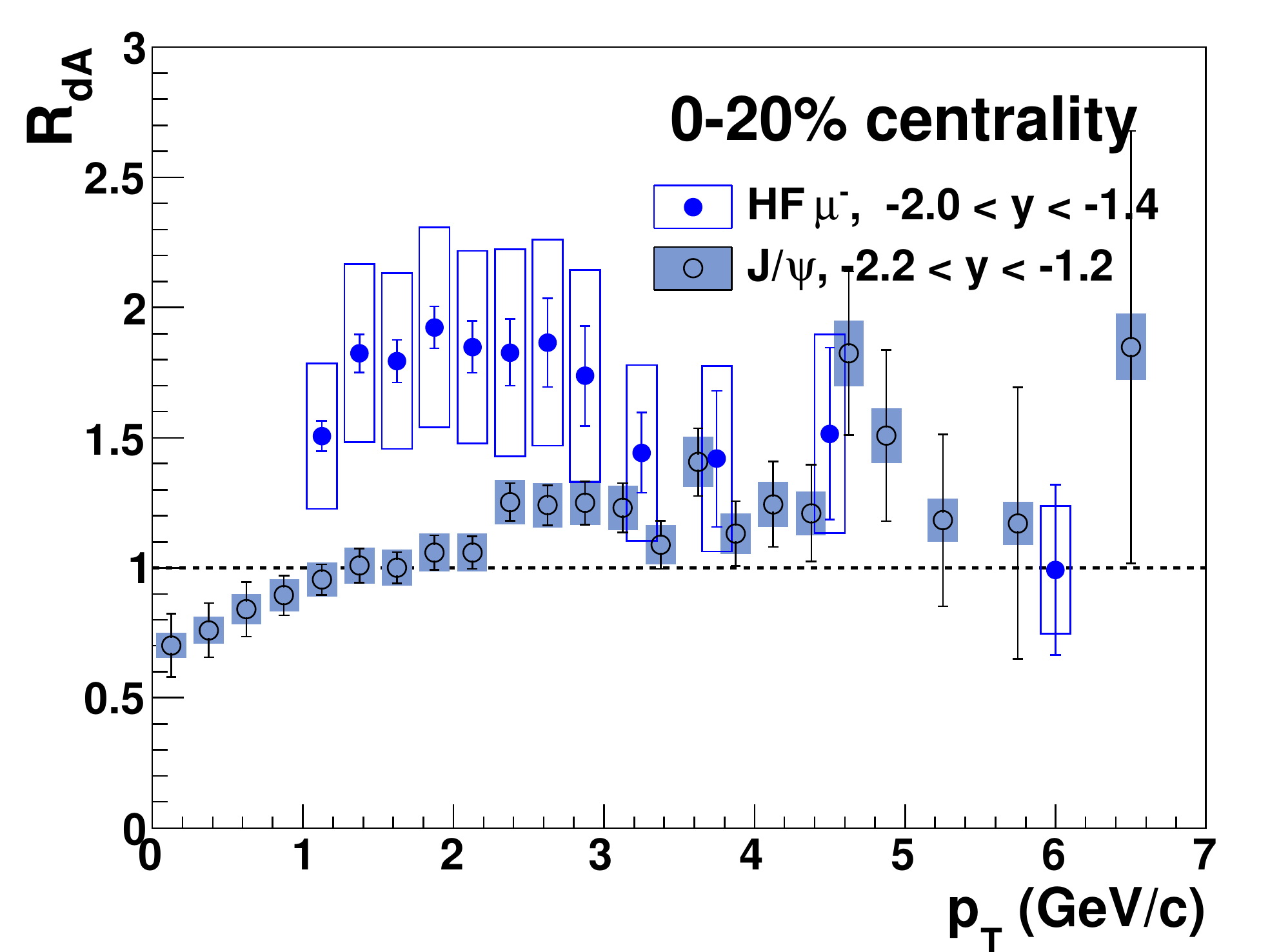}
\caption{$R_{dA}$ of open and hidden heavy flavor at backward rapidity (Au-going direction).}
\label{fig:bkwd}
\end{minipage}
\end{figure}

Charm production in a nuclear target is sensitive to a variety of initial and final state effects that do not occur in the more elementary $p+p$ system.  Since the dominant heavy quark production mechanism is gluon fusion, modifications to the gluon parton distribution function in the nucleus will affect the charm cross section.  Interactions with nuclear material can cause charm quarks to lose energy and/or broaden the charm $p_{T}$ spectrum.  Bound $c\bar{c}$ states have additional suppression mechanisms, since they can be broken up through interactions with the nucleus or with comoving particles. 

Measurements at RHIC have indeed confirmed that $J/\psi$ meson production is suppressed with respect to the open charm baseline.  Figs. \ref{fig:fwd} through \ref{fig:bkwd} show the nuclear modification factor $R_{dA}$ for $J/\psi$ mesons and leptons from open heavy flavor decays, at forward, mid-, and backward rapidities, respectively, using data from \cite{PPG125, PPG131, PPG153}.  At forward rapidity, both $J/\psi$ and open heavy flavor exhibit suppression compared to a $p+p$ reference.  However, at mid- and backward rapidity, open heavy flavor is enhanced while $J/\psi$ production is suppressed, indicating that there is an additional mechanism inhibiting charmonia formation.  We note here that care must be taken when making direct comparisons between the $p_{T}$ dependence of these measurements, due to the kinematic differences between fully reconstructed $J/\psi$ mesons and leptons from decays of heavy flavor hadrons.  

\begin{figure}[htbp]
\centering
\begin{minipage}{.45\textwidth}
\centering
\includegraphics[scale=0.18]{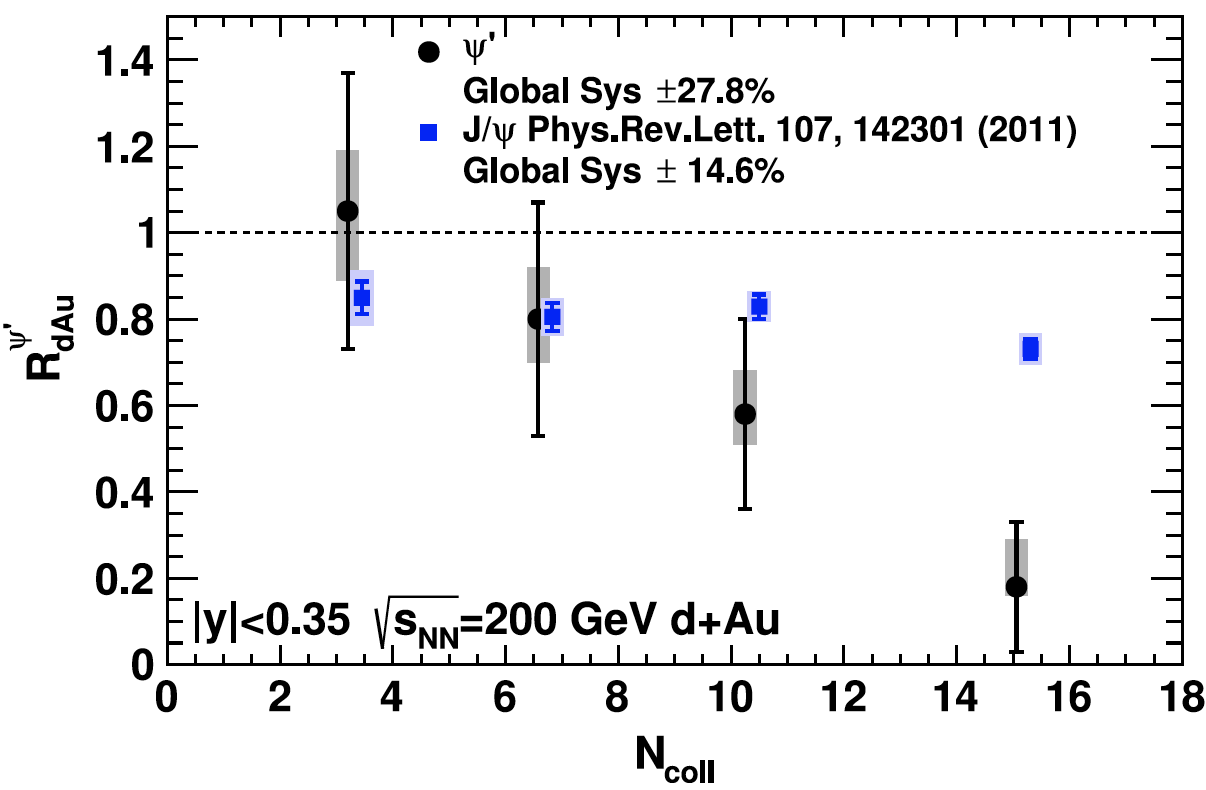}
\caption{The nuclear modification factor $R_{dA}$ for $J/\psi$ and $\psi(2s)$ production at midrapidity.}
\label{fig:RdA}
\end{minipage}\hfill
\begin{minipage}{.45\textwidth}
\centering
\includegraphics[scale=0.18]{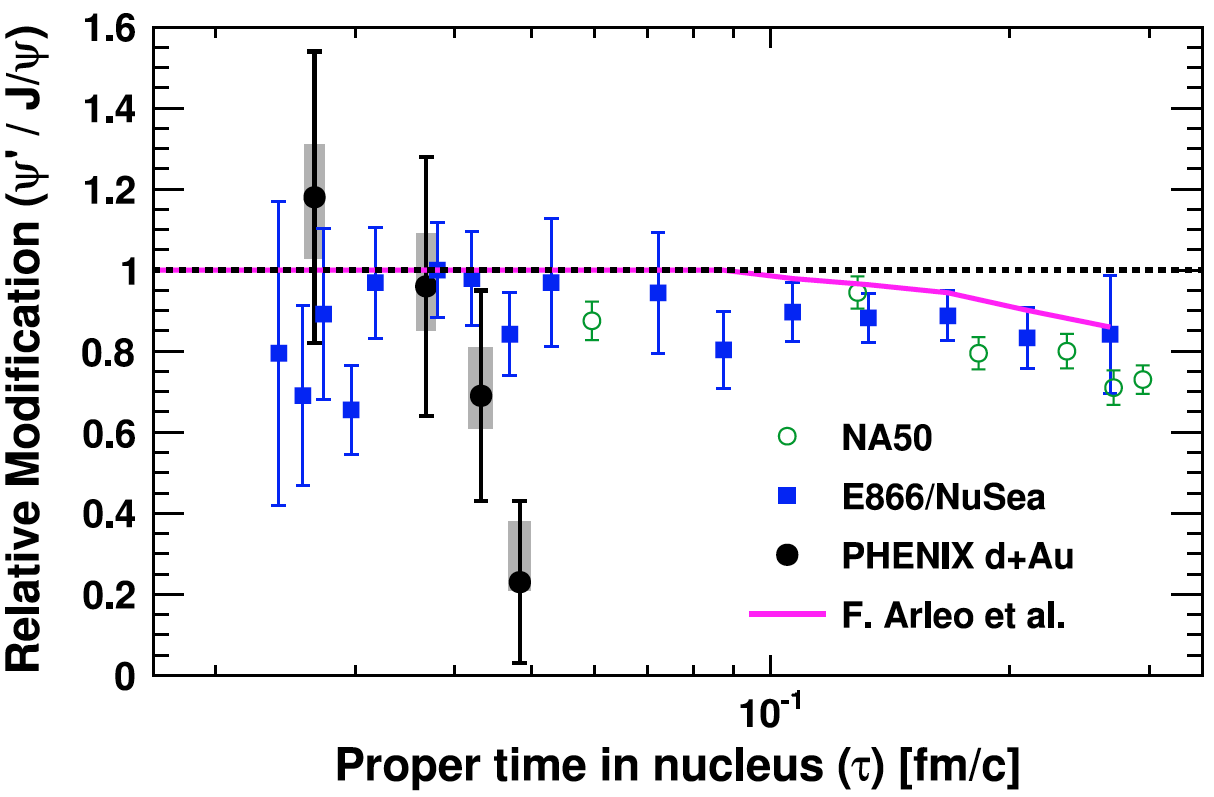}
\caption{Relative modification of $\psi(2s)$ and $J/\psi$ as a function of proper time spent in the nucleus, for PHENIX, NA50, and E866 \cite{PPG151, NA50_psip, E866_psip}.  A model which describes the fixed target data breaks down for the PHENIX data \cite{psip_time}.}
\label{fig:time}
\end{minipage}
\end{figure}

Several models that include various combinations of gluon shadowing, charm energy loss, multiple scattering in the nucleus, and nuclear break-up have been able to reproduce facets of the $J/\psi$ data \cite{Kopel,ferr,Vitev_Jpsi}.  As a further constraint, we turn to the excited charmonium state $\psi(2s)$.  Fig. \ref{fig:RdA} shows the nuclear modification factor $R_{dA}$ for $J/\psi$ and $\psi(2s)$, as a function of $N_{coll}$.  We observe that the $\psi(2s)$ has a markedly different trend than the $J/\psi$, and is more suppressed by a factor of $\sim$3 in the most central $d$+Au collisions.  

As with all charm, $\psi(2s)$ production is sensitive to initial state effects such as gluon shadowing and energy loss in the nucleus, and as a $c\bar{c}$ bound state it is subject to breakup effects.  Due to the short crossing time at RHIC, the average proper time the $c \bar{c}$ pair spends in the nucleus is only $\sim$0.05 fm/$c$, much shorter than the $J/\psi$ formation time of $\sim$0.15 fm/$c$ \cite{psip_time}. Since the charm quarks exist only as a precursor state while in the nucleus, no breakup mechanism that acts within that time scale can have different effects on $J/\psi$ and $\psi(2s)$ production.  However, effects which occur after hadronization may very well be more significant for the $\psi(2s)$, as its small binding energy of $\sim$50 MeV may allow it to be more easily disrupted.  

We see in Fig. \ref{fig:time} a model which successfully describes the $\psi(2s)$ modification relative to that of the $J/\psi$ in fixed target experiments \cite{psip_time}.  This model considers the precursor $c\bar{c}$ state as two quarks that expand linearly with time.  Due to its larger radius, the $\psi(2s)$ is more susceptible to breakup while in the nucleus, since it is effectively a larger target.  However, this model fails to describe the relative modification seen at PHENIX, where crossing times are very short.

These observations indicate that the mechanism behind the larger relative suppression of $\psi(2s)$ acts after hadronization, which at RHIC occurs outside the nucleus.  Interactions with comoving particles produced in the event could produce such an effect.

\section{New capabilities at Forward Rapidity}

The midrapidity data on $\psi(2s)$ suppression raises questions about the influence of comoving particles on charmonium outside the nucleus.  Measurements at forward and backward rapidity in $p$+Au collisions allow us to examine charmonium production under different conditions:  in the $p-$going direction, the nuclear crossing time is shorter and the produced particle density is relatively low.  In the Au-going direction, the time the precursor state spends in the nucleus is longer and the co-moving particle density is higher.  However, the existing PHENIX muon spectrometer arms do not have sufficient mass resolution to separate the $\psi(2s)$ peak from the much more prominent $J/\psi$ peak in the dimuon continuum.

\begin{figure}[htbp]
\centering
\begin{minipage}{.45\textwidth}
\centering
\includegraphics[scale=0.35]{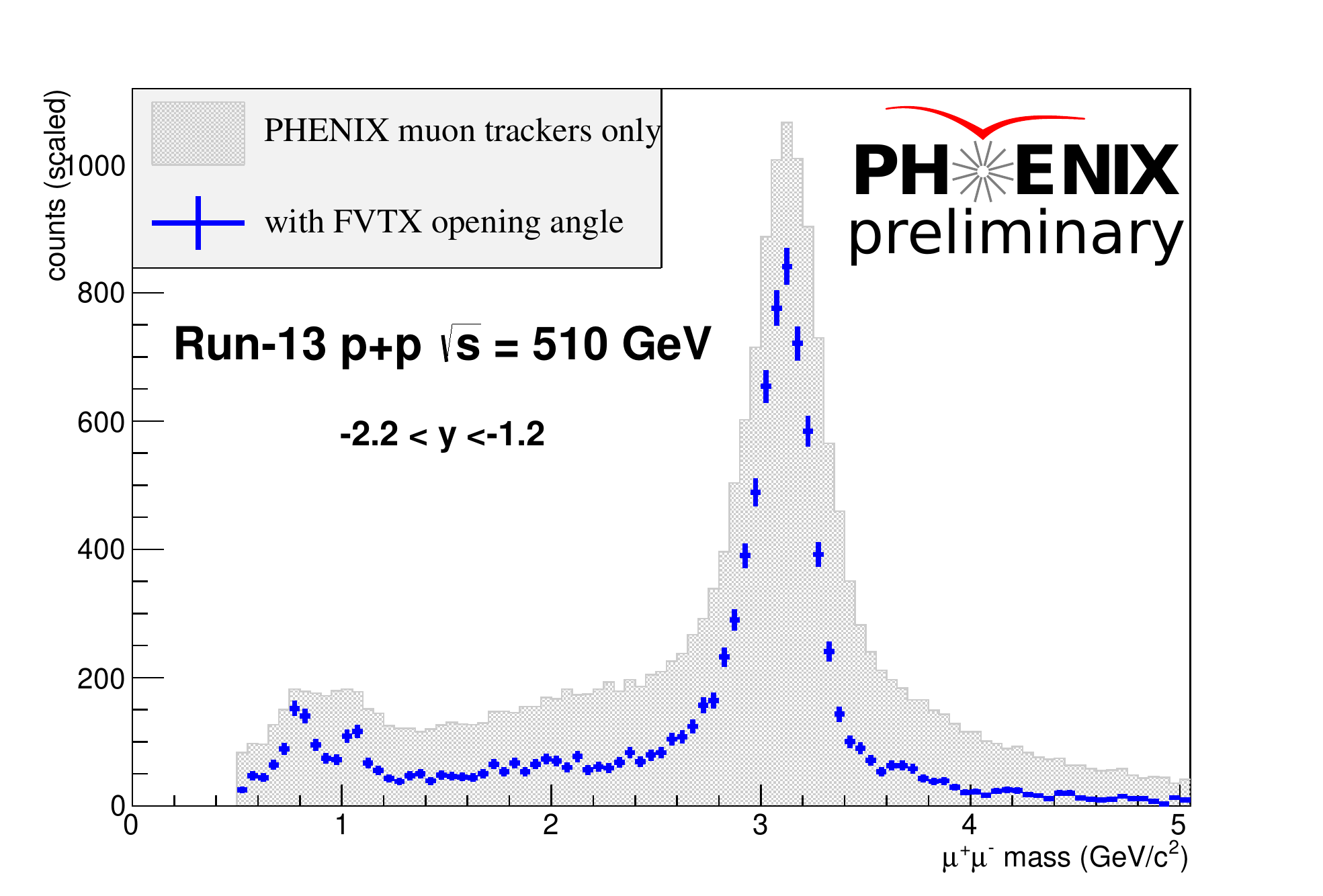}
\caption{Dimuon mass spectra in the PHENIX South muon arm using only the muon trackers, and the muon trackers with the FVTX.}
\label{fig:prelim_mass_comparison}
\end{minipage}\hfill
\begin{minipage}{.45\textwidth}
\centering
\includegraphics[scale=0.32]{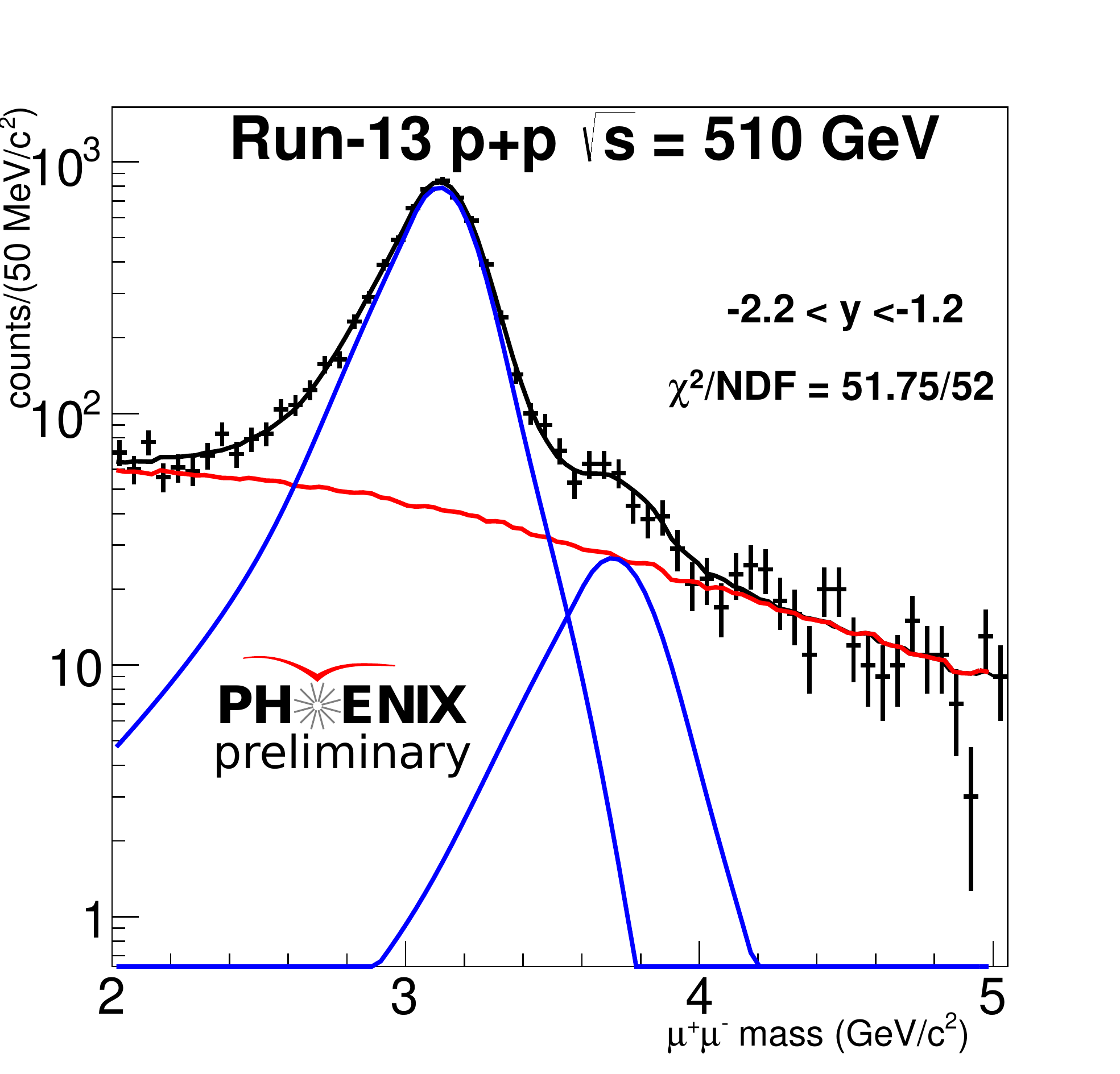}
\caption{Fit to dimuon mass spectra used to extract the $J/\psi$ and $\psi(2s)$ peaks.}
\label{fig:prelim_fit}
\end{minipage}
\end{figure}

The capability to separate the two states was achieved in 2012 with a major upgrade to the PHENIX muon spectrometers.  The Forward Silicon Vertex Detector (FVTX, \cite{FVTXNIM}) provides precise measurements of charged particle tracks in front of the hadron absorbers, allowing for a determination of muon pair opening angles before they undergo any multiple scattering in absorber material, which greatly improves the dimuon mass resolution.  Fig. \ref{fig:prelim_mass_comparison} shows the dimuon mass spectrum from the 2013 $p+p$ dataset using the pair opening angle as measured by the South PHENIX muon trackers (which are behind the absorber) and the FVTX.

\begin{figure}
\centering
\includegraphics[scale=0.5]{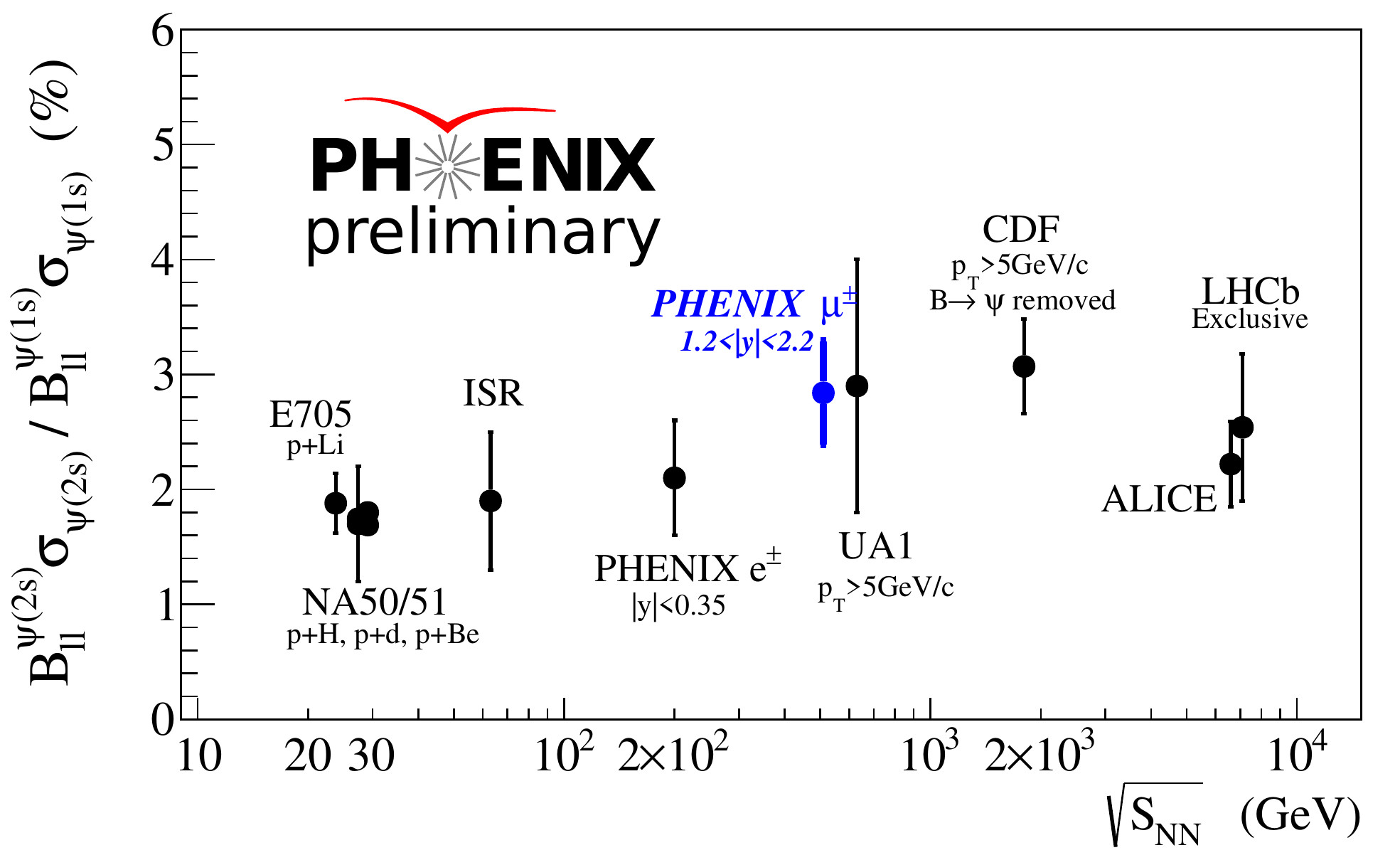}
\caption{World data on the ratio of $\psi(2s) / J/\psi$ production as a function of collision energy, in $p+p$ and $p$+A collisions.}
\label{fig:prelim_rat}
\end{figure}

The same data is shown in Fig. \ref{fig:prelim_fit}, with a reduced scale to focus on the mass region of interest.  A fit consisting of a Crystal Ball plus Gaussian is used to extract the $J/\psi$ and $\psi(2s)$ counts, with the background represented by the mixed-event combinatorial contribution plus an exponential.  During fitting, the difference between the peak centers is set to the PDG value of 0.589 GeV.  The ratio of the yields of these two states is corrected for the difference in detector efficiency, and shown in Fig. \ref{fig:prelim_rat}, along with world data from other experiments (see \cite{PPG104, ALICE_psip, LHCb_psip} and references therein).  The error bar on the ratio is the quadrature sum of the statistical uncertainty and the systematic uncertainty due to uncertainties in the relative widths of the Gaussian component of the fit, dimuon trigger efficiency, background contributions, and detector efficiency.  The new PHENIX measurement is consistent with world data and has an error bar that is comparable to other precision measurements.

\section{Conclusions and Outlook}

The midrapidity data on $\psi(2s)$ production in $d$+Au collisions at PHENIX indicates that there is an additional suppression mechanism that affects the 2s state more significantly than the $J/\psi$.  The short crossing time at RHIC suggests that this suppression occurs outside the nucleus, possibly through interactions with comoving particles.  With upgraded capabilities at forward and backward rapidity and proven results from the 2013 $p+p$ data, the PHENIX experiment is well prepared to measure $\psi(2s)$ production in an expanded kinematic range in the 2014 Au+Au and 2015 $p$+Au datasets.





\bibliographystyle{elsarticle-num}
\bibliography{Durham_QM14}

\begin{thebibliography}{10}
\expandafter\ifx\csname url\endcsname\relax
  \def\url#1{\texttt{#1}}\fi
\expandafter\ifx\csname urlprefix\endcsname\relax\def\urlprefix{URL }\fi
\expandafter\ifx\csname href\endcsname\relax
  \def\href#1#2{#2} \def\path#1{#1}\fi

\bibitem{Jpsi_theory_review}
N.~Brambilla, et~al., EPJ C 71~(2) (2011) 1--178.

\bibitem{MatsuiSatz}
T.~Matsui, H.~Satz, Phys. Lett. B 178~(4) (1986) 416 -- 422.

\bibitem{PHENIXNIM}
K.~Adcox, et~al., Nucl. Instrum. Meth. A 499 (2003) 469--479.

\bibitem{PPG151}
A.~Adare, et~al., Phys. Rev. Lett. 111 (2013) 202301.

\bibitem{PPG125}
A.~Adare, et~al., Phys. Rev. C 87 (2013) 034904.

\bibitem{PPG131}
A.~Adare, et~al., Phys. Rev. Lett. 112 (2014) 252301.

\bibitem{PPG153}
A.~Adare, et~al., Phys. Rev. Lett. 109 (2012) 242301.

\bibitem{Kopel}
B.~Kopeliovich, I.~Potashnikova, I.~Schmidt, Nucl. Phys. A 864~(1) (2011) 203
  -- 212.

\bibitem{ferr}
E.~Ferreiro, F.~Fleuret, J.~Lansberg, N.~Matagne, A.~Rakotozafindrabe, Few-Body
  Syst. 53~(1-2) (2012) 27--36.

\bibitem{Vitev_Jpsi}
R.~Sharma, I.~Vitev, Phys. Rev. C 87 (2013) 044905.

\bibitem{psip_time}
F.~Arleo, P.-B. Gossiaux, T.~Gousset, J.~Aichelin, Phys. Rev. C 61 (2000)
  054906.

\bibitem{NA50_psip}
M.~Abreu, et~al., Phys. Lett. B 449~(1–2) (1999) 128 -- 136.

\bibitem{E866_psip}
M.~J. Leitch, et~al., Phys. Rev. Lett. 84 (2000) 3256--3260.

\bibitem{FVTXNIM}
C.~Aidala, et~al., Nucl. Instrum. Meth. A 755 (2014) 44--61.

\bibitem{PPG104}
A.~Adare, et~al., Phys. Rev. D 85 (2012) 092004.

\bibitem{ALICE_psip}
B.~Abelev, et~al., arXiv:1403.3648.

\bibitem{LHCb_psip}
R.~Aaij, et~al., J. Phys. G 40~(4) (2013) 045001.

\end{thebibliography}

\end{document}